\documentclass[a4paper,11pt]{article}
\usepackage{pos}
\usepackage{pdfpages}

\title{Performance update of an event-type based analysis for the Cherenkov Telescope Array}
 \ShortTitle{Event types performance update for CTA}

\author*[a]{J. Bernete}
\author[b]{O. Gueta}
\author[a]{T. Hassan}
\author[c]{M. Linhoff}
\author[b]{G. Maier}
\author[d]{A. Sinha}

\affiliation[a]{Centro de Investigaciones Energéticas, Medioambientales y Tecnológicas (CIEMAT), Av. Complutense, 40, 28040 Madrid, Spain}

\affiliation[b]{DESY, Platanenallee 6, 15738 Zeuthen, Germany}

\affiliation[c]{Department of Physics, TU Dortmund University, Otto-Hahn-Str. 4a, 44227 Dortmund, Germany}

\affiliation[d]{IPARCOS-UCM, Instituto de Física de Partículas y del Cosmos, and EMFTEL Department, Universidad Complutense de Madrid, E-28040 Madrid, Spain}

\onbehalf{for the CTA Consortium} 

\emailAdd{juan.bernete@ciemat.es}

\abstract{The Cherenkov Telescope Array (CTA) will be the next-generation observatory in the field of very-high-energy (20 GeV to 300 TeV) gamma-ray astroparticle physics. The traditional approach to data analysis in this field is to apply quality cuts, optimized using Monte Carlo simulations, on the data acquired to maximize sensitivity. Subsequent steps of the analysis typically use the surviving events to calculate one set of instrument response functions (IRFs) to physically interpret the results. However, an alternative approach is the use of event types, as implemented in experiments such as the \textit{Fermi}-LAT. This approach divides events into sub-samples based on their reconstruction quality, and a set of IRFs is calculated for each sub-sample. The sub-samples are then combined in a joint analysis, treating them as independent observations. In previous works we demonstrated that event types, classified using Machine Learning methods according to their expected angular reconstruction quality, have the potential to significantly improve the CTA angular and energy resolution of a point-like source analysis. Now, we validated the production of event-type wise full-enclosure IRFs, ready to be used with science tools (such as \textit{Gammapy} and \textit{ctools}). We will report on the impact of using such an event-type classification on CTA high-level performance, compared to the traditional procedure.}

\ConferenceLogo{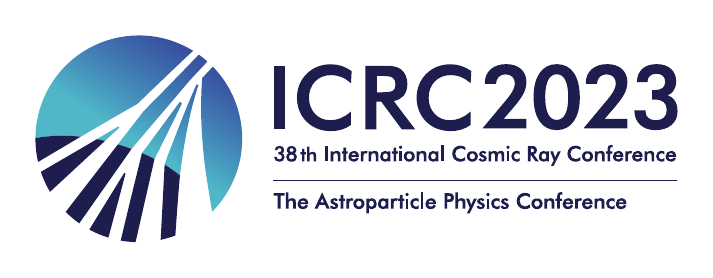}

\FullConference{%
38th International Cosmic Ray Conference (ICRC2023)\\
  26 July - 3 August, 2023\\
  Nagoya, Japan}

\begin{document}
\maketitle

\section{Introduction}

The Cherenkov Telescope Array (CTA)\footnote{www.cta-observatory.org} represents the next-generation observatory in the field of very-high-energy gamma-ray astroparticle physics. It employs two arrays of imaging atmospheric Cherenkov telescopes (IACTs), one for each hemisphere, composed of telescopes of three different sizes. Its  optimized configuration provides a major improvement in sensitivity and in angular and energy resolution with respect to the current generation of IACTs over a very broad energy range from 20 GeV up to more than 300 TeV.

The performance of this future observatory is estimated from detailed Monte Carlo (MC) simulations, described by a set of Instrument Response Functions (IRFs). The main IRF components describing the instrument performance to gamma-ray observations are the effective area, the energy dispersion and point-spread function (PSF). These IRFs are then used by science tools (such as gammapy \cite{gammapy:2017} and ctools \cite{ctools}) to simulate the instrument performance over specific science cases. The methodology to calculate the expected sensitivity and associated IRFs of CTA, as well as their detailed description, has been described in previous contributions (see \cite{Bernl_hr_2013, Acharyya_2019, HASSAN201776}) and is briefly discussed in section \ref{IRF}.

The \textit{Fermi} Large Area Telescope (LAT) Collaboration \cite{Atwood_2009} proved that high-level analysis performance can be significantly improved by separating events for which the response of the detector is different into event types and producing specific IRFs for each event type \cite{bruel2018fermilat}. By including this extra knowledge into the likelihood analysis, multiple benefits are achieved: reducing background contamination, increasing the effective area and sensitivity as well as significantly improving the angular and energy resolution for a subset of the events. Inspired by the success of event types in \textit{Fermi}-LAT, we present in this work the status of an analog implementation for IACTs, specifically for the future CTA.

This work is a natural continuation of Ref. \cite{Hassan_2021}, where we demonstrated that event types are able to improve the angular and energy resolution by up to 25\% for a point-like source located at the center of the field of view (FoV). This first step did not allow the generalized use of event-type-wise IRFs at the science tools level to properly evaluate their impact over specific science cases. 

In this work, we have validated the production of event-type wise offset-dependent point-like and full-enclosure IRFs for CTA (i.e. valid for both point-like or extended sources located anywhere within the FoV). These IRFs, tailored to each event type, are now ready to be used by science tools. We also present the impact of this event-type classification on the high-level performance of CTA, comparing it to the standard procedure (not using event types), as well as evaluate the potential for further improvement with a better event-type classification. 

\section{Event type partitioning}\label{ETpar}

Previous work successfully demonstrated the effectiveness of machine learning (ML) methods in separating event types based on their expected quality in angular reconstruction \cite{Hassan_2021}. Our approach begins at the Data Level 2 (DL2), as the product of a classical IACT analysis, which classification score called \textit{gammaness} and a list of lower-level parameters describing individual telescope images and stereo parameters (such as Hillas parameterization, reconstructed altitude of shower maximum, etc...).

An event type is a unique tag for each event in a DL2 table that classifies of all them in terms of their quality in angular reconstruction. We use a ML regression model to predict the angular difference between true and reconstructed direction (from now on, predicted misdirection), so the division of event types reduces to establishing thresholds for the top X\% reconstructed events (lowest predicted misdirection), the following Y\%, etc. Where the number of event types and their proportions can be freely decided.

The event type partitioning methodology employed for this study is almost identical to the one described in the previous contribution \cite{Hassan_2021} with the following differences:

\begin{itemize}
    \item The MC simulated data used is diffuse (covering the full FoV of CTA telescopes) for gammas, protons and electrons.
    \item The regression model we use, a multilayer perceptron (MLP) neural network with a $tanh$ as neuron activation function, has been further optimized.
    \item The thresholds in predicted misdirection to divide the event types are now dependent in both energy and offset angle, instead of only energy.
\end{itemize}

\section{IRF production}\label{IRF}

The standard methodology to compute CTA IRFs \cite{Bernl_hr_2013, Acharyya_2019, HASSAN201776} starts from DL2 table. A re-weight of the simulated events is needed so that they resemble the particle statistics expected from a CTA observation of a Crab-Nebula-like source (as a test case). To compute IRFs a cut optimization is needed, generally  maximizing sensitivity as a function of the reconstructed energy. Events surviving these quality cuts are the ones that will be used to compute the final set of IRFs. The cut optimization is usually performed over the following parameters: multiplicity (number of telescopes used in the reconstruction of an event), \textit{gammaness} and, in the case of a point-like source analysis, the angular size of the signal region (\textit{ON region}). Once CTA data is produced, the list of events surviving the \textit{gammaness} and multiplicity cuts together with their corresponding IRFs form the Data Level 3 (DL3) products.

With this procedure, the amount of data surviving quality cuts (and therefore actually used in the analysis) is small compared to the rejected data, while the latter could still be useful. Furthermore, as there is only one set of IRFs generated applied equally for all  events, all the extra knowledge we have from the low-level analysis is lost.

In an event-type based analysis, the event type partitioning (as explained in Section \ref{ETpar}) occurs before optimizing the cuts and computing the IRFs. This allows to create a number of independent lists (as many as event types), each one with their corresponding set of IRFs describing their average quality.

\begin{figure}
    \centering
    \includegraphics[width=0.9\textwidth]{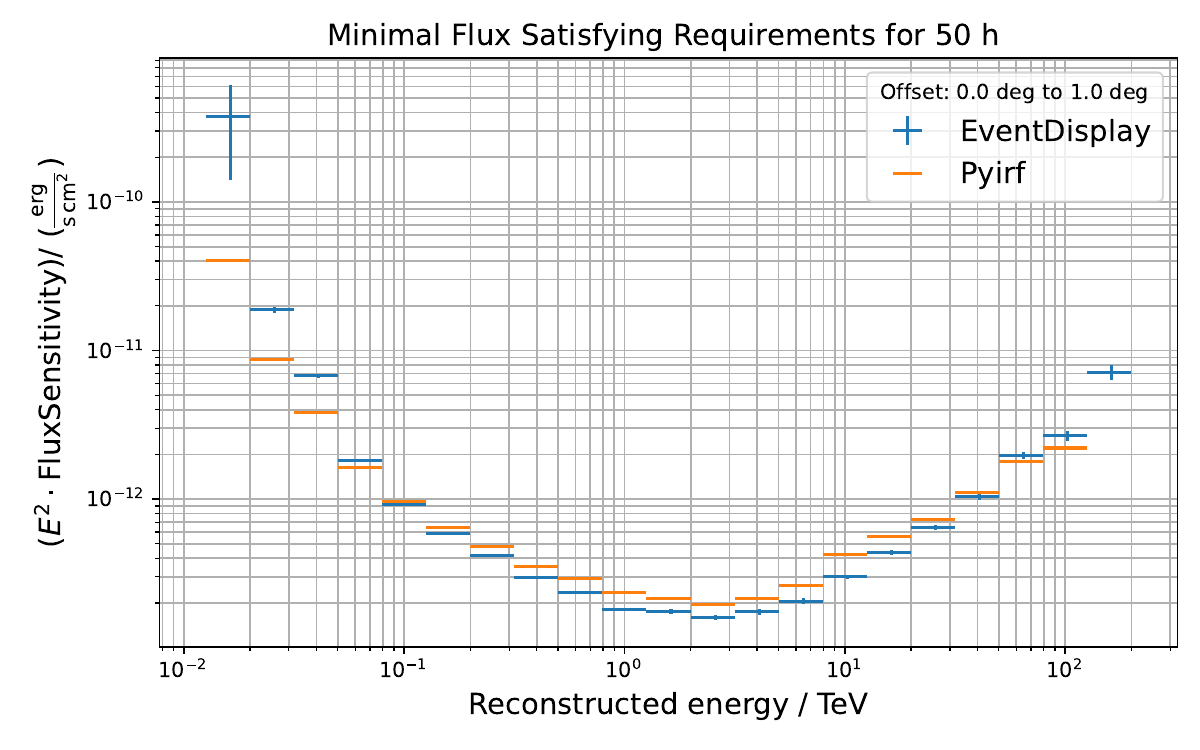}
    \caption{\textit{EventDisplay} and \textit{pyirf} comparison of the resulting sensitivity for a Crab-like observation of 50 hours in the central FoV offset bin.}
    \label{fig:comparison}
\end{figure}

To compute the IRFs and store them in the proper format\footnote{https://gamma-astro-data-formats.readthedocs.io/en/v0.3/irfs/index.html} we used the library \textit{pyirf}\footnote{https://github.com/cta-observatory/pyirf}. This library first needed to be tested and validated to produce offset-dependent and full-enclosure IRFs. To validate it, we compared the resulting sensitivity and IRFs to the ones computed by \textit{EventDisplay} \cite{maier2017eventdisplay} with the same MC data. The tests consisted in two steps:

\begin{enumerate}
    \item Validate pyirf IRF computation. By using identical DL2 tables, we compared the computed IRF components by using exactly the same quality cuts as \textit{EventDisplay}. The results were identical, and therefore the computation of all IRF components was validated.
    \item Validate pyirf cut optimization. We performed two independent cut optimizations (with \textit{pyirf} and \textit{EventDisplay}) by selecting the cuts that provide a better sensitivity in each energy bin, and compared resulting sensitivities. As shown in Fig. \ref{fig:comparison}, they are not exactly the same but they agree to within 50\% between 30 GeV and 100 TeV (also across different values of the FoV). The reason of the disagreement is not known, but is probably related to small differences in the cut selection methods (for example, \textit{EventDisplay} uses smaller bins for the direction cuts).
\end{enumerate}

After performing these tests, we conclude \textit{pyirf} is suitable to our needs, as it allows us to compute both point-like and full-enclosure IRFs properly for all different camera offset angles up to $6$ $deg$. Once the production of IRFs was validated, we produced various sets of event-type-wise IRFs, ready to be used with high-level science tools, in this case, \textit{Gammapy}.

\section{Results}

We evaluate the expected angular reconstruction quality of  all events, rank them and eventually classify them into different event-type partitionings to then produce event-type-wise offset-dependent IRFs for 50 hours of observing time for the "Alpha" layout of CTA-North (4 LSTs and 9 MSTs) \cite{Gueta_2021}.

\begin{figure}
    \centering
    \includegraphics[width=\textwidth]{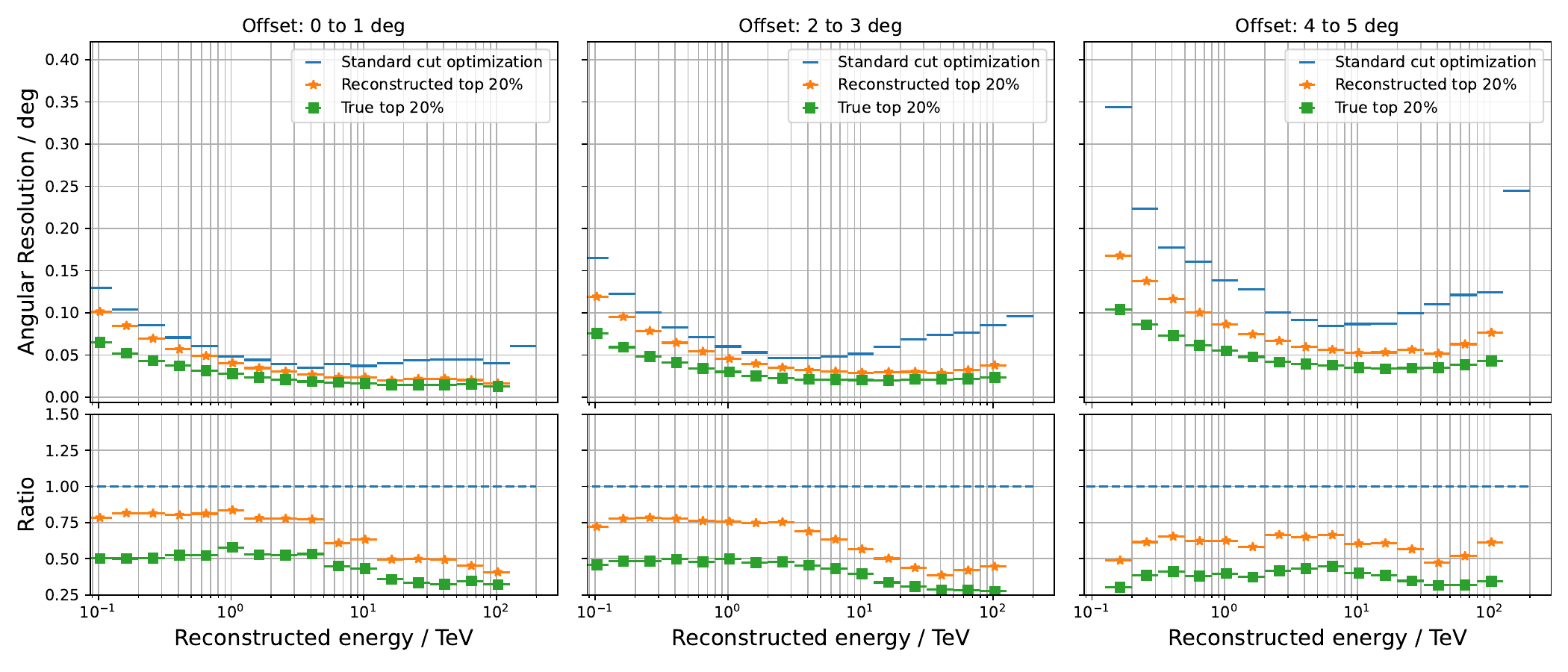}
    \caption{Angular resolution for a 50 hours observation, comparison between the standard cuts case, the reconstructed top 20\% events and the true top 20\%. Repeated for different offset ranges.}
    \label{fig:angular}
\end{figure}

By computing the angular resolution for the ranked top 20\% events as reconstructed by our model, we show a 25 to 50 \% improvement in angular resolution with respect to the standard cut optimization method (not using event types), as shown in Figure \ref{fig:angular}. We also computed the angular resolution for the true top 20\% events, i.e.: ranking by the actual difference between the reconstructed and the true simulated position of each event, so we can see there is still room for improvement of our regression model.

\begin{figure}
    \centering
    \includegraphics[width=0.8\textwidth]{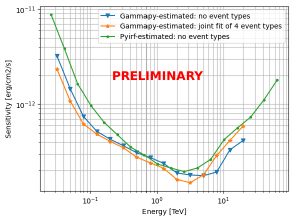}
    \caption{Preliminary sensitivity curve reconstructed with \textit{Gammapy} by doing a likelihood analysis with combined event types (4 types with 25\% of the events each) and with no event types, compared to the standard sensitivity computed with \textit{pyirf}. Note that Gammapy-estimated sensitivity does not take into account any conditions on background systematics and minimum number of excess events, which affect the highest and the lowest energies.}
    \label{fig:sensitivity}
\end{figure}

We can use these IRFs to perform either 1D (spectral evaluations of point-like sources) or 3D (spectral and morphological studies) simulations with \textit{Gammapy}. Datasets are simulated from a set of IRFs: we are able to perform simulations for a single IRFs set and for event-type-wise IRFs treating them as independent samples that may be combined in a joint-likelihood analysis. By doing this with a Crab-like source simulations over a wide range of fluxes, we can reconstruct the combined sensitivity from all event types as shown in Figure \ref{fig:sensitivity}, by identifying over each bin in reconstructed energy the simulated flux that provides a 5$\sigma$ detection. Note this method to compute sensitivity (for any set of observations or simulations at \textit{Gammapy} level) does not have the usual requirements generally included in the calculation of sensitivity, such as the requirement of the excess being larger than a 5\% of the background (to account for systematics in the background) or the minimum number of 10 excess events (heavily affecting the sensitivity at the highest energies), which is the main reason of the disagreements at the lowest and highest energies with the \textit{pyirf}-estimated curve.

\section{Conclusions}

The conclusions of this work can be summarized by the following milestones:

\begin{enumerate}
    \item Our ML regression model is able to predict the misdirection of each event and, therefore, can be used to separate event types. It should be noted there is still room for improvement.
    \item Offset-dependence has been introduced and validated in the event-type partitioning process.
    \item We are now able to produce consistently both point-like and full-enclosure event-type-wise IRFs over the full FoV, which allows high-level simulations with science tools such as \textit{Gammapy}.
    \item Event-type-wise IRFs show a significant improvement in angular resolution (25 to 50\% over a subset of the events).
    \item Preliminary \textit{Gammapy} analysis already shows that is possible to combine observations from different event-type samples for a better performance.
\end{enumerate}

This work shows the great potential that an event-type based analysis could have for improving CTA's performance. A specific science case for fundamental physics with gamma-ray propagation \cite{Abdalla_2021} that could be benefited by event types is measuring intergalactic magnetic fields, in which the size of the PSF is crucial. Another important example is the Galactic Plane Survey, where the improved angular resolution at large offset angles will allow to separate sources and determine extensions and morphologies better than ever in this energy range.

\section*{Acknowledgements}

This work was conducted in the context of the CTA Consortium and CTA Observatory.
We gratefully acknowledge financial support from the agencies and organizations listed here: http://www.cta-observatory.org/consortium\_acknowledgments.

\bibliographystyle{plain}
\bibliography{refs.bib}

%% Full authors list (ONLY FOR COLLABORATIONS)
%\clearpage
%\section*{Full Authors List: \Coll\ Collaboration}
%
%\noindent \textbf{Note comment afterwards:} Collaborations have the possibility to provide an authors list in xml format which will be used while generating the DOI entries making the full authors list searchable in databases like Inspire HEP. For instructions please go to icrc2021.desy.de/proceedings or contact us under icrc2021proc@desy.de.\\
%
%\scriptsize
%\noindent
%first.author$^1$, 
%second.author$^2$, 
%third.author$^3$ % .... more names
%and 
%last.author$^{n}$ \\
%
%\noindent
%$^1$first.affiliation.
%$^2$second.affiliation. % .... more affiliation
%$^{m}$last.affiliation.



\section*{Supplementary material}

# The CTA Consortium

K. Abe[1], S. Abe[2], A. Acharyya[3], R. Adam[4,5], A. Aguasca-Cabot[6], I. Agudo[7], J. Alfaro[8],
N. Alvarez-Crespo[9], R. Alves Batista[10], J.-P. Amans[11], E. Amato[12], F. Ambrosino[13],
E. O. Angüner[14], L. A. Antonelli[13], C. Aramo[15], C. Arcaro[16], L. Arrabito[17], K. Asano[2],
J. Aschersleben[18], H. Ashkar[5], L. Augusto Stuani[19], D. Baack[20], M. Backes[21,22],
C. Balazs[23], M. Balbo[24], A. Baquero Larriva[9,25], V. Barbosa Martins[26],
U. Barres de Almeida[27,28], J. A. Barrio[9], D. Bastieri[29], P. I. Batista[26], I. Batkovic[29],
R. Batzofin[30], J. Baxter[2], G. Beck[31], J. Becker Tjus[32], L. Beiske[20], D. Belardinelli[33],
W. Benbow[34], E. Bernardini[29], J. Bernete Medrano[35], K. Bernlöhr[36], A. Berti[37],
V. Beshley[38], P. Bhattacharjee[39], S. Bhattacharyya[40], B. Bi[41], N. Biederbeck[20],
A. Biland[42], E. Bissaldi[43,44], O. Blanch[45], J. Blazek[46], C. Boisson[11], J. Bolmont[47],
G. Bonnoli[48,49], P. Bordas[6], Z. Bosnjak[50], F. Bradascio[51], C. Braiding[52], E. Bronzini[53],
R. Brose[54], A. M. Brown[55], F. Brun[51], G. Brunelli[53,7], A. Bulgarelli[53], I. Burelli[56],
L. Burmistrov[57], M. Burton[58,59], T. Bylund[60], P. G. Calisse[61], A. Campoy-Ordaz[62],
B. K. Cantlay[63,64], M. Capalbi[65], A. Caproni[66], R. Capuzzo-Dolcetta[13], C. Carlile[67],
S. Caroff[39], A. Carosi[13], R. Carosi[49], M.-S. Carrasco[68], E. Cascone[69], F. Cassol[68],
N. Castrejon[70], F. Catalani[71], D. Cerasole[72], M. Cerruti[73], S. Chaty[73], A. W. Chen[31],
M. Chernyakova[74], A. Chiavassa[75,76], J. Chudoba[46], C. H. Coimbra Araujo[77],
V. Conforti[53], F. Conte[36], J. L. Contreras[9], C. Cossou[60], A. Costa[78], H. Costantini[68],
P. Cristofari[11], O. Cuevas[79], Z. Curtis-Ginsberg[80], G. D'Amico[81], F. D'Ammando[82],
M. Dadina[53], M. Dalchenko[57], L. David[26], I. D. Davids[21], F. Dazzi[83], A. De Angelis[29],
M. de Bony de Lavergne[60], V. De Caprio[69], G. De Cesare[53], E. M. de Gouveia Dal Pino[28],
B. De Lotto[56], M. De Lucia[15], R. de Menezes[75,76], M. de Naurois[5],
E. de Ona Wilhelmi[26], N. De Simone[26], V. de Souza[19], L. del Peral[70], M. V. del Valle[28],
E. Delagnes[84], A. G. Delgado Giler[19,18], C. Delgado[35], M. Dell'aiera[39], R. Della Ceca[48],
M. Della Valle[69], D. della Volpe[57], D. Depaoli[36], A. Dettlaff[37], T. Di Girolamo[85,15],
A. Di Piano[53], F. Di Pierro[75], R. Di Tria[72], L. Di Venere[44], C. Díaz-Bahamondes[8],
C. Dib[86], S. Diebold[41], R. Dima[29], A. Dinesh[9], A. Djannati-Ataï[73], J. Djuvsland[81],
A. Domínguez[9], R. M. Dominik[20], A. Donini[13], D. Dorner[87,42], J. Dörner[32], M. Doro[29],
R. D. C. dos Anjos[77], J.-L. Dournaux[11], D. Dravins[67], C. Duangchan[88,64], C. Dubos[89],
L. Ducci[41], V. V. Dwarkadas[90], J. Ebr[46], C. Eckner[39,91], K. Egberts[30], S. Einecke[52],
D. Elsässer[20], G. Emery[68], M. Escobar Godoy[92], J. Escudero[7], P. Esposito[93,94],
D. Falceta-Goncalves[95], V. Fallah Ramazani[32], A. Faure[17], E. Fedorova[13,96], S. Fegan[5],
K. Feijen[73], Q. Feng[34], G. Ferrand[97,98], F. Ferrarotto[99], E. Fiandrini[100], A. Fiasson[39],
V. Fioretti[53], L. Foffano[101], L. Font Guiteras[62], G. Fontaine[5], S. Fröse[20], S. Fukami[42],
Y. Fukui[102], S. Funk[88], D. Gaggero[49], G. Galanti[94], G. Galaz[8], Y. A. Gallant[17],
S. Gallozzi[13], V. Gammaldi[10], C. Gasbarra[33], M. Gaug[62], A. Ghalumyan[103],
F. Gianotti[53], M. Giarrusso[104], N. Giglietto[43,44], F. Giordano[72], A. Giuliani[94],
J.-F. Glicenstein[51], J. Glombitza[88], P. Goldoni[105], J. M. González[106], M. M. González[107],
J. Goulart Coelho[108], J. Granot[109,110], D. Grasso[49], R. Grau[45], D. Green[37], J. G. Green[37],
T. Greenshaw[111], G. Grolleron[47], J. Grube[112], O. Gueta[26], S. Gunji[113], D. Hadasch[2],
P. Hamal[46], W. Hanlon[34], S. Hara[114], V. M. Harvey[52], K. Hashiyama[2], T. Hassan[35],
M. Heller[57], S. Hernández Cadena[107], J. Hie[115], N. Hiroshima[2], B. Hnatyk[96], R. Hnatyk[96],
D. Hoffmann[68], W. Hofmann[36], M. Holler[116], D. Horan[5], P. Horvath[117], T. Hovatta[118],
D. Hrupec[119], S. Hussain[28,120], M. Iarlori[121], T. Inada[2], F. Incardona[78], Y. Inome[2],
S. Inoue[98], F. Iocco[85,15], K. Ishio[122], M. Jamrozy[123], P. Janecek[46], F. Jankowsky[124],
C. Jarnot[115], P. Jean[115], I. Jiménez Martínez[35], W. Jin[3], L. Jocou[125], C. Juramy-Gilles[47],
J. Jurysek[46], O. Kalekin[88], D. Kantzas[91], V. Karas[126], S. Kaufmann[55], D. Kerszberg[45],
B. Khélifi[73], D. B. Kieda[127], T. Kleiner[26], W. Kluźniak[128], Y. Kobayashi[2], K. Kohri[129],
N. Komin[31], P. Kornecki[11], K. Kosack[60], H. Kubo[2], J. Kushida[1], A. La Barbera[65],
N. La Palombara[94], M. Láinez[9], A. Lamastra[13], J. Lapington[130], S. Lazarević[131],
J. Lazendic-Galloway[23], S. Leach[130], M. Lemoine-Goumard[132], J.-P. Lenain[47], G. Leto[78],
F. Leuschner[41], E. Lindfors[118], M. Linhoff[20], I. Liodakis[118], L. Loïc[51], S. Lombardi[13],
F. Longo[133], R. López-Coto[7], M. López-Moya[9], A. López-Oramas[134], S. Loporchio[43,44],
J. Lozano Bahilo[70], P. L. Luque-Escamilla[135], O. Macias[136], G. Maier[26], P. Majumdar[137],
D. Malyshev[41], D. Malyshev[88], D. Mandat[46], G. Manicò[104,138], P. Marinos[52],
S. Markoff[136], I. Márquez[7], P. Marquez[45], G. Marsella[139,104], J. Martí[135], P. Martin[115],

G. A. Martínez[35], M. Martínez[45], O. Martinez[140,141], C. Marty[115], A. Mas-Aguilar[9],
M. Mastropietro[13], G. Maurin[39], W. Max-Moerbeck[142], D. Mazin[2,37], D. Melkumyan[26],
S. Menchiari[12,49], E. Mestre[143], J.-L. Meunier[47], D. M.-A. Meyer[30], D. Miceli[16],
M. Michailidis[41], J. Michałowski[144], T. Miener[9], J. M. Miranda[140,145], A. Mitchell[88],
M. Mizote[146], T. Mizuno[147], R. Moderski[128], L. Mohrmann[36], M. Molero[134], C. Molfese[83],
E. Molina[134], T. Montaruli[57], A. Moralejo[45], D. Morcuende[9,7], K. Morik[20], A. Morselli[33],
E. Moulin[51], V. Moya Zamanillo[9], R. Mukherjee[148], K. Munari[78], A. Muraczewski[128],
H. Muraishi[149], T. Nakamori[113], L. Nava[48], A. Nayak[55], R. Nemmen[28,150], L. Nickel[20],
J. Niemiec[144], D. Nieto[9], M. Nievas Rosillo[134], M. Nikołajuk[151], K. Nishijima[1],
K. Noda[2], D. Nosek[152], B. Novosyadlyj[153], V. Novotny[152], S. Nozaki[37], P. O'Brien[130],
M. Ohishi[2], Y. Ohtani[2], A. Okumura[154,155], J.-F. Olive[115], B. Olmi[156,12], R. A. Ong[157],
M. Orienti[82], R. Orito[158], M. Orlandini[53], E. Orlando[133], M. Ostrowski[123], N. Otte[159],
I. Oya[61], I. Pagano[78], A. Pagliaro[65], M. Palatiello[56], G. Panebianco[53], J. M. Paredes[6],
N. Parmiggiani[53], S. R. Patel[89], B. Patricelli[13,160], D. Pavlović[161], A. Pe'er[37], M. Pech[46],
M. Pecimotika[161,162], M. Peresano[76,75], J. Pérez-Romero[10,40], G. Peron[73], M. Persic[163,164],
P.-O. Petrucci[125], O. Petruk[38], F. Pfeifle[87], F. Pintore[65], G. Pirola[37], C. Pittori[13],
C. Plard[39], F. Podobnik[165], M. Pohl[30,26], E. Pons[39], E. Prandini[29], J. Prast[39],
G. Principe[133], C. Priyadarshi[45], N. Produit[24], D. Prokhorov[136], E. Pueschel[26],
G. Pühlhofer[41], M. L. Pumo[138,104], M. Punch[73], A. Quirrenbach[124], S. Rainò[72],
N. Randazzo[104], R. Rando[29], T. Ravel[115], S. Razzaque[166,110], M. Regeard[73],
P. Reichherzer[167,32], A. Reimer[116], O. Reimer[116], A. Reisenegger[8,168], T. Reposeur[132],
B. Reville[36], W. Rhode[20], M. Ribó[6], T. Richtler[169], F. Rieger[36], E. Roache[34],
G. Rodriguez Fernandez[33], M. D. Rodríguez Frías[70], J. J. Rodríguez-Vázquez[35], P. Romano[48],
G. Romeo[78], J. Rosado[9], G. Rowell[52], B. Rudak[128], A. J. Ruiter[170], C. B. Rulten[55],
F. Russo[53], I. Sadeh[26], L. Saha[34], T. Saito[2], S. Sakurai[2], H. Salzmann[41], D. Sanchez[39],
M. Sánchez-Conde[10], P. Sangiorgi[65], H. Sano[2], M. Santander[3], A. Santangelo[41],
R. Santos-Lima[28], A. Sanuy[6], T. Šarić[171], A. Sarkar[26], S. Sarkar[167], F. G. Saturni[13],
V. Savchenko[172], A. Scherer[8], P. Schipani[69], B. Schleicher[87,42], P. Schovanek[46],
J. L. Schubert[20], F. Schussler[51], U. Schwanke[173], G. Schwefer[36], S. Scuderi[94],
M. Seglar Arroyo[45], I. Seitenzahl[170], O. Sergijenko[96,174,175], V. Sguera[53], R. Y. Shang[157],
P. Sharma[89], G. D. S. SIDIBE[84], L. Sidoli[94], H. Siejkowski[176], C. Siqueira[19], P. Sizun[84],
V. Sliusar[24], A. Slowikowska[177], H. Sol[11], A. Specovius[88], S. T. Spencer[88,167],
D. Spiga[48], A. Stamerra[13,178], S. Stanič[40], T. Starecki[179], R. Starling[130], C. Steppa[30],
T. Stolarczyk[60], J. Strišković[119], M. Strzys[2], Y. Suda[180], T. Suomijärvi[89], D. Tak[26],
M. Takahashi[154], R. Takeishi[2], P.-H. T. Tam[2,181], S. J. Tanaka[182], T. Tanaka[146],
K. Terauchi[183], V. Testa[13], L. Tibaldo[115], O. Tibolla[55], F. Torradeflot[184,35], D. F. Torres[143],
E. Torresi[53], N. Tothill[131], F. Toussenel[47], V. Touzard[115], A. Tramacere[24], P. Travnicek[46],
G. Tripodo[139,104], S. Truzzi[165], A. Tsiahina[115], A. Tutone[65], M. Vacula[117,46], B. Vallage[51],
P. Vallania[75,185], R. Vallés[143], C. van Eldik[88], J. van Scherpenberg[37], J. Vandenbroucke[80],
V. Vassiliev[157], P. Venault[84], S. Ventura[165], S. Vercellone[48], G. Verna[165], A. Viana[19],
N. Viaux[186], A. Vigliano[56], J. Vignatti[86], C. F. Vigorito[75,76], V. Vitale[33], V. Vodeb[40],
V. Voisin[47], S. Vorobiov[40], G. Voutsinas[57], I. Vovk[2], V. Waegebaert[115], S. J. Wagner[124],
R. Walter[24], M. Ward[55], M. Wechakama[63,64], R. White[36], A. Wierzcholska[144], M. Will[37],
D. A. Williams[92], F. Wohlleben[36], A. Wolter[48], T. Yamamoto[146], R. Yamazaki[182],
L. Yang[166,181], T. Yoshida[187], T. Yoshikoshi[2], M. Zacharias[124,22], R. Zanmar Sanchez[78],
D. Zavrtanik[40], M. Zavrtanik[40], A. A. Zdziarski[128], A. Zech[11], V. I. Zhdanov[96],
K. Ziętara[123], M. Živec[40], J. Zuriaga-Puig[10]

# Affiliations

[1] Department of Physics, Tokai University, 4-1-1, Kita-Kaname, Hiratsuka, Kanagawa 259-1292, Japan

[2] Institute for Cosmic Ray Research, University of Tokyo, 5-1-5, Kashiwa-no-ha, Kashiwa, Chiba 277-8582, Japan

[3] University of Alabama, Tuscaloosa, Department of Physics and Astronomy, Gallalee Hall, Box 870324 Tuscaloosa, AL 35487-0324, USA

[4] Université Côte d'Azur, Observatoire de la Côte d'Azur, CNRS, Laboratoire Lagrange, France

[5] Laboratoire Leprince-Ringuet, CNRS/IN2P3, École polytechnique, Institut Polytechnique de Paris, 91120 Palaiseau, France

[6] Departament de Física Quàntica i Astrofísica, Institut de Ciències del Cosmos, Universitat de Barcelona, IEEC-UB, Martí i Franquès, 1, 08028, Barcelona, Spain

[7] Instituto de Astrofísica de Andalucía-CSIC, Glorieta de la Astronomía s/n, 18008, Granada, Spain

[8] Pontificia Universidad Católica de Chile, Av. Libertador Bernardo O'Higgins 340, Santiago, Chile

[9] IPARCOS-UCM, Instituto de Física de Partículas y del Cosmos, and EMFTEL Department, Universidad Complutense de Madrid, E-28040 Madrid, Spain

[10] Instituto de Física Teórica UAM/CSIC and Departamento de Física Teórica, Universidad Autónoma de Madrid, c/ Nicolás Cabrera 13-15, Campus de Cantoblanco UAM, 28049 Madrid, Spain

[11] LUTH, GEPI and LERMA, Observatoire de Paris, Université PSL, Université Paris Cité, CNRS, 5 place Jules Janssen, 92190, Meudon, France

[12] INAF - Osservatorio Astrofisico di Arcetri, Largo E. Fermi, 5 - 50125 Firenze, Italy

[13] INAF - Osservatorio Astronomico di Roma, Via di Frascati 33, 00040, Monteporzio Catone, Italy

[14] TÜBİTAK Research Institute for Fundamental Sciences, 41470 Gebze, Kocaeli, Turkey

[15] INFN Sezione di Napoli, Via Cintia, ed. G, 80126 Napoli, Italy

[16] INFN Sezione di Padova, Via Marzolo 8, 35131 Padova, Italy

[17] Laboratoire Univers et Particules de Montpellier, Université de Montpellier, CNRS/IN2P3, CC 72, Place Eugène Bataillon, F-34095 Montpellier Cedex 5, France

[18] Kapteyn Astronomical Institute, University of Groningen, Landleven 12, 9747 AD, Groningen, The Netherlands

[19] Instituto de Física de São Carlos, Universidade de São Paulo, Av. Trabalhador São-carlense, 400 - CEP 13566-590, São Carlos, SP, Brazil

[20] Astroparticle Physics, Department of Physics, TU Dortmund University, Otto-Hahn-Str. 4a, 44227 Dortmund, Germany

[21] Department of Physics, Chemistry & Material Science, University of Namibia, Private Bag 13301, Windhoek, Namibia

[22] Centre for Space Research, North-West University, Potchefstroom, 2520, South Africa

[23] School of Physics and Astronomy, Monash University, Melbourne, Victoria 3800, Australia

[24] Department of Astronomy, University of Geneva, Chemin d'Ecogia 16, CH-1290 Versoix, Switzerland

[25] Faculty of Science and Technology, Universidad del Azuay, Cuenca, Ecuador.

[26] Deutsches Elektronen-Synchrotron, Platanenallee 6, 15738 Zeuthen, Germany

[27] Centro Brasileiro de Pesquisas Físicas, Rua Xavier Sigaud 150, RJ 22290-180, Rio de Janeiro, Brazil

[28] Instituto de Astronomia, Geofísica e Ciências Atmosféricas - Universidade de São Paulo, Cidade Universitária, R. do Matão, 1226, CEP 05508-090, São Paulo, SP, Brazil

[29] INFN Sezione di Padova and Università degli Studi di Padova, Via Marzolo 8, 35131 Padova, Italy

[30] Institut für Physik & Astronomie, Universität Potsdam, Karl-Liebknecht-Strasse 24/25, 14476 Potsdam, Germany

[31] University of the Witwatersrand, 1 Jan Smuts Avenue, Braamfontein, 2000 Johannesburg, South Africa

[32] Institut für Theoretische Physik, Lehrstuhl IV: Plasma-Astroteilchenphysik, Ruhr-Universität Bochum, Universitätsstraße 150, 44801 Bochum, Germany

[33] INFN Sezione di Roma Tor Vergata, Via della Ricerca Scientifica 1, 00133 Rome, Italy

[34] Center for Astrophysics | Harvard & Smithsonian, 60 Garden St, Cambridge, MA 02138, USA

[35] CIEMAT, Avda. Complutense 40, 28040 Madrid, Spain

[36] Max-Planck-Institut für Kernphysik, Saupfercheckweg 1, 69117 Heidelberg, Germany

[37] Max-Planck-Institut für Physik, Föhringer Ring 6, 80805 München, Germany

[38] Pidstryhach Institute for Applied Problems in Mechanics and Mathematics NASU, 3B Naukova Street, Lviv, 79060, Ukraine

[39] Univ. Savoie Mont Blanc, CNRS, Laboratoire d'Annecy de Physique des Particules - IN2P3, 74000 Annecy, France

[40] Center for Astrophysics and Cosmology (CAC), University of Nova Gorica, Nova Gorica, Slovenia

[41] Institut für Astronomie und Astrophysik, Universität Tübingen, Sand 1, 72076 Tübingen, Germany

[42] ETH Zürich, Institute for Particle Physics and Astrophysics, Otto-Stern-Weg 5, 8093 Zürich, Switzerland

[43] Politecnico di Bari, via Orabona 4, 70124 Bari, Italy

[44] INFN Sezione di Bari, via Orabona 4, 70126 Bari, Italy

[45] Institut de Fisica d'Altes Energies (IFAE), The Barcelona Institute of Science and Technology, Campus UAB, 08193 Bellaterra (Barcelona), Spain

[46] FZU - Institute of Physics of the Czech Academy of Sciences, Na Slovance 1999/2, 182 21 Praha 8, Czech Republic

[47] Sorbonne Université, CNRS/IN2P3, Laboratoire de Physique Nucléaire et de Hautes Energies, LPNHE, 4 place Jussieu, 75005 Paris, France

[48] INAF - Osservatorio Astronomico di Brera, Via Brera 28, 20121 Milano, Italy

[49] INFN Sezione di Pisa, Edificio C – Polo Fibonacci, Largo Bruno Pontecorvo 3, 56127 Pisa

[50] University of Zagreb, Faculty of electrical engineering and computing, Unska 3, 10000 Zagreb, Croatia

[51] IRFU, CEA, Université Paris-Saclay, Bât 141, 91191 Gif-sur-Yvette, France

[52] School of Physics, Chemistry and Earth Sciences, University of Adelaide, Adelaide SA 5005, Australia

[53] INAF - Osservatorio di Astrofisica e Scienza dello spazio di Bologna, Via Piero Gobetti 93/3, 40129 Bologna, Italy

[54] Dublin Institute for Advanced Studies, 31 Fitzwilliam Place, Dublin 2, Ireland

[55] Centre for Advanced Instrumentation, Department of Physics, Durham University, South Road, Durham, DH1 3LE, United Kingdom

[56] INFN Sezione di Trieste and Università degli Studi di Udine, Via delle Scienze 208, 33100 Udine, Italy

[57] University of Geneva - Département de physique nucléaire et corpusculaire, 24 rue du Général-Dufour, 1211 Genève 4, Switzerland

[58] Armagh Observatory and Planetarium, College Hill, Armagh BT61 9DG, United Kingdom

[59] School of Physics, University of New South Wales, Sydney NSW 2052, Australia

[60] Université Paris-Saclay, Université Paris Cité, CEA, CNRS, AIM, F-91191 Gif-sur-Yvette Cedex, France

[61] Cherenkov Telescope Array Observatory, Saupfercheckweg 1, 69117 Heidelberg, Germany

[62] Unitat de Física de les Radiacions, Departament de Física, and CERES-IEEC, Universitat Autònoma de Barcelona, Edifici C3, Campus UAB, 08193 Bellaterra, Spain

[63] Department of Physics, Faculty of Science, Kasetsart University, 50 Ngam Wong Wan Rd., Lat Yao, Chatuchak, Bangkok, 10900, Thailand

[64] National Astronomical Research Institute of Thailand, 191 Huay Kaew Rd., Suthep, Muang, Chiang Mai, 50200, Thailand

[65] INAF - Istituto di Astrofisica Spaziale e Fisica Cosmica di Palermo, Via U. La Malfa 153, 90146 Palermo, Italy

[66] Universidade Cruzeiro do Sul, Núcleo de Astrofísica Teórica (NAT/UCS), Rua Galvão Bueno 8687, Bloco B, sala 16, Libertade 01506-000 - São Paulo, Brazil

[67] Lund Observatory, Lund University, Box 43, SE-22100 Lund, Sweden

[68] Aix Marseille Univ, CNRS/IN2P3, CPPM, Marseille, France

[69] INAF - Osservatorio Astronomico di Capodimonte, Via Salita Moiariello 16, 80131 Napoli, Italy

[70] Universidad de Alcalá - Space & Astroparticle group, Facultad de Ciencias, Campus Universitario Ctra. Madrid-Barcelona, Km. 33.600 28871 Alcalá de Henares (Madrid), Spain

[71] Escola de Engenharia de Lorena, Universidade de São Paulo, Área I - Estrada Municipal do Campinho, s/n°, CEP 12602-810, Pte. Nova, Lorena, Brazil

[72] INFN Sezione di Bari and Università degli Studi di Bari, via Orabona 4, 70124 Bari, Italy

[73] Université Paris Cité, CNRS, Astroparticule et Cosmologie, F-75013 Paris, France

[74] Dublin City University, Glasnevin, Dublin 9, Ireland

[75] INFN Sezione di Torino, Via P. Giuria 1, 10125 Torino, Italy

[76] Dipartimento di Fisica - Università degli Studi di Torino, Via Pietro Giuria 1 - 10125 Torino, Italy

[77] Universidade Federal Do Paraná - Setor Palotina, Departamento de Engenharias e Exatas, Rua Pioneiro, 2153, Jardim Dallas, CEP: 85950-000 Palotina, Paraná, Brazil

[78] INAF - Osservatorio Astrofisico di Catania, Via S. Sofia, 78, 95123 Catania, Italy

[79] Universidad de Valparaíso, Blanco 951, Valparaiso, Chile

[80] University of Wisconsin, Madison, 500 Lincoln Drive, Madison, WI, 53706, USA

[81] Department of Physics and Technology, University of Bergen, Museplass 1, 5007 Bergen, Norway

[82] INAF - Istituto di Radioastronomia, Via Gobetti 101, 40129 Bologna, Italy

[83] INAF - Istituto Nazionale di Astrofisica, Viale del Parco Mellini 84, 00136 Rome, Italy

[84] IRFU/DEDIP, CEA, Université Paris-Saclay, Bat 141, 91191 Gif-sur-Yvette, France

[85] Universitá degli Studi di Napoli "Federico II" - Dipartimento di Fisica "E. Pancini", Complesso universitario di Monte Sant'Angelo, Via Cintia - 80126 Napoli, Italy

[86] CCTVal, Universidad Técnica Federico Santa María, Avenida España 1680, Valparaíso, Chile

[87] Institute for Theoretical Physics and Astrophysics, Universität Würzburg, Campus Hubland Nord, Emil-Fischer-Str. 31, 97074 Würzburg, Germany

[88] Friedrich-Alexander-Universität Erlangen-Nürnberg, Erlangen Centre for Astroparticle Physics, Nikolaus-Fiebiger-Str. 2, 91058 Erlangen, Germany

[89] Université Paris-Saclay, CNRS/IN2P3, IJCLab, 91405 Orsay, France

[90] Department of Astronomy and Astrophysics, University of Chicago, 5640 S Ellis Ave, Chicago, Illinois, 60637, USA

[91] LAPTh, CNRS, USMB, F-74940 Annecy, France

[92] Santa Cruz Institute for Particle Physics and Department of Physics, University of California, Santa Cruz, 1156 High Street, Santa Cruz, CA 95064, USA

[93] University School for Advanced Studies IUSS Pavia, Palazzo del Broletto, Piazza della Vittoria 15, 27100 Pavia, Italy

[94] INAF - Istituto di Astrofisica Spaziale e Fisica Cosmica di Milano, Via A. Corti 12, 20133 Milano, Italy

[95] Escola de Artes, Ciências e Humanidades, Universidade de São Paulo, Rua Arlindo Bettio, CEP 03828-000, 1000 São Paulo, Brazil

[96] Astronomical Observatory of Taras Shevchenko National University of Kyiv, 3 Observatorna Street, Kyiv, 04053, Ukraine

[97] The University of Manitoba, Dept of Physics and Astronomy, Winnipeg, Manitoba R3T 2N2, Canada

[98] RIKEN, Institute of Physical and Chemical Research, 2-1 Hirosawa, Wako, Saitama, 351-0198, Japan

[99] INFN Sezione di Roma La Sapienza, P.le Aldo Moro, 2 - 00185 Roma, Italy

[100] INFN Sezione di Perugia and Università degli Studi di Perugia, Via A. Pascoli, 06123 Perugia, Italy

[101] INAF - Istituto di Astrofisica e Planetologia Spaziali (IAPS), Via del Fosso del Cavaliere 100, 00133 Roma, Italy

[102] Department of Physics, Nagoya University, Chikusa-ku, Nagoya, 464-8602, Japan

[103] Alikhanyan National Science Laboratory, Yerevan Physics Institute, 2 Alikhanyan Brothers St., 0036, Yerevan, Armenia

[104] INFN Sezione di Catania, Via S. Sofia 64, 95123 Catania, Italy

[105] Université Paris Cité, CNRS, CEA, Astroparticule et Cosmologie, F-75013 Paris, France

[106] Universidad Andres Bello, República 252, Santiago, Chile

[107] Universidad Nacional Autónoma de México, Delegación Coyoacán, 04510 Ciudad de México, Mexico

[108] Núcleo de Astrofísica e Cosmologia (Cosmo-ufes) & Departamento de Física, Universidade Federal do Espírito Santo (UFES), Av. Fernando Ferrari, 514. 29065-910. Vitória-ES, Brazil

[109] Astrophysics Research Center of the Open University (ARCO), The Open University of Israel, P.O. Box 808, Ra'anana 4353701, Israel

[110] Department of Physics, The George Washington University, Washington, DC 20052, USA

[111] University of Liverpool, Oliver Lodge Laboratory, Liverpool L69 7ZE, United Kingdom

[112] King's College London, Strand, London, WC2R 2LS, United Kingdom

[113] Department of Physics, Yamagata University, Yamagata, Yamagata 990-8560, Japan

[114] Learning and Education Development Center, Yamanashi-Gakuin University, Kofu, Yamanashi 400-8575, Japan

[115] IRAP, Université de Toulouse, CNRS, CNES, UPS, 9 avenue Colonel Roche, 31028 Toulouse, Cedex 4, France

[116] Universität Innsbruck, Institut für Astro- und Teilchenphysik, Technikerstr. 25/8, 6020 Innsbruck, Austria

[117] Palacký University Olomouc, Faculty of Science, Joint Laboratory of Optics of Palacký University and Institute of Physics of the Czech Academy of Sciences, 17. listopadu 1192/12, 779 00 Olomouc, Czech Republic

[118] Finnish Centre for Astronomy with ESO, University of Turku, Finland, FI-20014 University of Turku, Finland

[119] Josip Juraj Strossmayer University of Osijek, Trg Ljudevita Gaja 6, 31000 Osijek, Croatia

[120] Gran Sasso Science Institute (GSSI), Viale Francesco Crispi 7, 67100 L'Aquila, Italy and INFN-Laboratori Nazionali del Gran Sasso (LNGS), via G. Acitelli 22, 67100 Assergi (AQ), Italy

[121] Dipartimento di Scienze Fisiche e Chimiche, Università degli Studi dell'Aquila and GSGC-LNGS-INFN, Via Vetoio 1, L'Aquila, 67100, Italy

[122] Faculty of Physics and Applied Computer Science, University of Łódź, ul. Pomorska 149-153, 90-236 Łódź, Poland

[123] Astronomical Observatory, Jagiellonian University, ul. Orla 171, 30-244 Cracow, Poland

[124] Landessternwarte, Zentrum für Astronomie der Universität Heidelberg, Königstuhl 12, 69117 Heidelberg, Germany

[125] Univ. Grenoble Alpes, CNRS, IPAG, 414 rue de la Piscine, Domaine Universitaire, 38041 Grenoble Cedex 9, France

[126] Astronomical Institute of the Czech Academy of Sciences, Bocni II 1401 - 14100 Prague, Czech Republic

[127] Department of Physics and Astronomy, University of Utah, Salt Lake City, UT 84112-0830, USA

[128] Nicolaus Copernicus Astronomical Center, Polish Academy of Sciences, ul. Bartycka 18, 00-716 Warsaw, Poland

[129] Institute of Particle and Nuclear Studies, KEK (High Energy Accelerator Research Organization), 1-1 Oho, Tsukuba, 305-0801, Japan

[130] School of Physics and Astronomy, University of Leicester, Leicester, LE1 7RH, United Kingdom

[131] Western Sydney University, Locked Bag 1797, Penrith, NSW 2751, Australia

[132] Université Bordeaux, CNRS, LP2I Bordeaux, UMR 5797, 19 Chemin du Solarium, F-33170 Gradignan, France

[133] INFN Sezione di Trieste and Università degli Studi di Trieste, Via Valerio 2 I, 34127 Trieste, Italy

[134] Instituto de Astrofísica de Canarias and Departamento de Astrofísica, Universidad de La Laguna, La Laguna, Tenerife, Spain

[135] Escuela Politécnica Superior de Jaén, Universidad de Jaén, Campus Las Lagunillas s/n, Edif. A3, 23071 Jaén, Spain

[136] Anton Pannekoek Institute/GRAPPA, University of Amsterdam, Science Park 904 1098 XH Amsterdam, The Netherlands

[137] Saha Institute of Nuclear Physics, A CI of Homi Bhabha National Institute, Kolkata 700064, West Bengal, India

[138] Università degli studi di Catania, Dipartimento di Fisica e Astronomia "Ettore Majorana", Via S. Sofia 64, 95123 Catania, Italy

[139] Dipartimento di Fisica e Chimica "E. Segrè", Università degli Studi di Palermo, Via Archirafi 36, 90123, Palermo, Italy

[140] UCM-ELEC group, EMFTEL Department, University Complutense of Madrid, 28040 Madrid, Spain

[141] Departamento de Ingeniería Eléctrica, Universidad Pontificia de Comillas - ICAI, 28015 Madrid

[142] Universidad de Chile, Av. Libertador Bernardo O'Higgins 1058, Santiago, Chile

[143] Institute of Space Sciences (ICE, CSIC), and Institut d'Estudis Espacials de Catalunya (IEEC), and Institució Catalana de Recerca I Estudis Avançats (ICREA), Campus UAB, Carrer de Can Magrans, s/n 08193 Cerdanyola del Vallés, Spain

[144] The Henryk Niewodniczański Institute of Nuclear Physics, Polish Academy of Sciences, ul. Radzikowskiego 152, 31-342 Cracow, Poland

[145] IPARCOS Institute, Faculty of Physics (UCM), 28040 Madrid, Spain

[146] Department of Physics, Konan University, Kobe, Hyogo, 658-8501, Japan

[147] Hiroshima Astrophysical Science Center, Hiroshima University, Higashi-Hiroshima, Hiroshima 739-8526, Japan

[148] Department of Physics, Columbia University, 538 West 120th Street, New York, NY 10027, USA

[149] School of Allied Health Sciences, Kitasato University, Sagamihara, Kanagawa 228-8555, Japan

[150] Kavli Institute for Particle Astrophysics and Cosmology, Stanford University, Stanford, CA 94305, USA

[151] University of Białystok, Faculty of Physics, ul. K. Ciołkowskiego 1L, 15-245 Białystok, Poland

[152] Charles University, Institute of Particle & Nuclear Physics, V Holešovičkách 2, 180 00 Prague 8, Czech Republic

[153] Astronomical Observatory of Ivan Franko National University of Lviv, 8 Kyryla i Mephodia Street, Lviv, 79005, Ukraine

[154] Institute for Space—Earth Environmental Research, Nagoya University, Furo-cho, Chikusa-ku, Nagoya 464-8601, Japan

[155] Kobayashi—Maskawa Institute for the Origin of Particles and the Universe, Nagoya University, Furo-cho, Chikusa-ku, Nagoya 464-8602, Japan

[156] INAF - Osservatorio Astronomico di Palermo "G.S. Vaiana", Piazza del Parlamento 1, 90134 Palermo, Italy

[157] Department of Physics and Astronomy, University of California, Los Angeles, CA 90095, USA

[158] Graduate School of Technology, Industrial and Social Sciences, Tokushima University, Tokushima 770-8506, Japan

[159] School of Physics & Center for Relativistic Astrophysics, Georgia Institute of Technology, 837 State Street, Atlanta, Georgia, 30332-0430, USA

[160] University of Pisa, Largo B. Pontecorvo 3, 56127 Pisa, Italy

[161] University of Rijeka, Faculty of Physics, Radmile Matejcic 2, 51000 Rijeka, Croatia

[162] Rudjer Boskovic Institute, Bijenicka 54, 10 000 Zagreb, Croatia

[163] INAF - Osservatorio Astronomico di Padova, Vicolo dell'Osservatorio 5, 35122 Padova, Italy

[164] INAF - Osservatorio Astronomico di Padova and INFN Sezione di Trieste, gr. coll. Udine, Via delle Scienze 208 I-33100 Udine, Italy

[165] INFN and Università degli Studi di Siena, Dipartimento di Scienze Fisiche, della Terra e dell'Ambiente (DSFTA), Sezione di Fisica, Via Roma 56, 53100 Siena, Italy

[166] Centre for Astro-Particle Physics (CAPP) and Department of Physics, University of Johannesburg, PO Box 524, Auckland Park 2006, South Africa

[167] University of Oxford, Department of Physics, Clarendon Laboratory, Parks Road, Oxford, OX1 3PU, United Kingdom

[168] Departamento de Física, Facultad de Ciencias Básicas, Universidad Metropolitana de Ciencias de la Educación, Avenida José Pedro Alessandri 774, Ñuñoa, Santiago, Chile

[169] Departamento de Astronomía, Universidad de Concepción, Barrio Universitario S/N, Concepción, Chile

[170] University of New South Wales, School of Science, Australian Defence Force Academy, Canberra, ACT 2600, Australia

[171] University of Split - FESB, R. Boskovica 32, 21 000 Split, Croatia

[172] EPFL Laboratoire d'astrophysique, Observatoire de Sauverny, CH-1290 Versoix, Switzerland

[173] Department of Physics, Humboldt University Berlin, Newtonstr. 15, 12489 Berlin, Germany

[174] Main Astronomical Observatory of the National Academy of Sciences of Ukraine, Zabolotnoho str., 27, 03143, Kyiv, Ukraine

[175] Space Technology Centre, AGH University of Science and Technology, Aleja Mickiewicza, 30, 30-059, Kraków, Poland

[176] Academic Computer Centre CYFRONET AGH, ul. Nawojki 11, 30-950, Kraków, Poland

[177] Institute of Astronomy, Faculty of Physics, Astronomy and Informatics, Nicolaus Copernicus University in Toruń, ul. Grudziądzka 5, 87-100 Toruń, Poland

[178] Cherenkov Telescope Array Observatory gGmbH, Via Gobetti, Bologna, Italy

[179] Warsaw University of Technology, Faculty of Electronics and Information Technology, Institute of Electronic Systems, Nowowiejska 15/19, 00-665 Warsaw, Poland

[180] Physics Program, Graduate School of Advanced Science and Engineering, Hiroshima University, 739-8526 Hiroshima, Japan

[181] School of Physics and Astronomy, Sun Yat-sen University, Zhuhai, China

[182] Department of Physical Sciences, Aoyama Gakuin University, Fuchinobe, Sagamihara, Kanagawa, 252-5258, Japan

[183] Division of Physics and Astronomy, Graduate School of Science, Kyoto University, Sakyo-ku, Kyoto, 606-8502, Japan

[184] Port d'Informació Científica, Edifici D, Carrer de l'Albareda, 08193 Bellaterrra (Cerdanyola del Vallès), Spain

[185] INAF - Osservatorio Astrofisico di Torino, Strada Osservatorio 20, 10025 Pino Torinese (TO), Italy

[186] Departamento de Física, Universidad Técnica Federico Santa María, Avenida España, 1680 Valparaíso, Chile

[187] Faculty of Science, Ibaraki University, Mito, Ibaraki, 310-8512, Japan



\end{document}